\documentclass[nofootinbib,preprint,aps,draft,pre,superscriptaddress,citeautoscript,floatfix]{revtex4-1}

\setcitestyle{super}
\usepackage{graphics}
\usepackage{graphicx}
\usepackage{mathrsfs}
\usepackage{amsmath}
\usepackage{bm}
\usepackage{color}
\usepackage{float}

\begin{document}

\title{Swelling Dynamics of a Disk-Shaped Gel}
\author{Xingkun Man}
\email{manxk@buaa.edu.cn}
\affiliation{Center of Soft Matter Physics and its Applications, Beihang University, Beijing 100191, China}
\affiliation{School of Physics, Beihang University, Beijing 100191, China}
\author{Masao Doi}
\email{masao.doi@buaa.edu.cn}
\affiliation{Center of Soft Matter Physics and its Applications, Beihang University, Beijing 100191, China}
\affiliation{School of Physics, Beihang University, Beijing 100191, China}

\begin{abstract}

When a gel absorbs solvent from surrounding, stress field is created in the gel, and this causes complex dynamics of the swelling behavior.  Here we study this effect for disk-shaped gel by rigorously solving the diffusio-mechanical coupling equation. We show that (a) while the macroscopic thickness and the radius of the gel increases monotonically in time, the gel is compressed near the mid-plane, and that (b) while the swelling time depends on the shear modulus $G$ of the gel, its dependence is weak, and the time is mainly determined by the friction constant of the gel network and the osmotic bulk modulus of the gel. We also show that these characteristic features are reproduced accurately by a simple variational calculation for the gel deformation. An analytical expression is given for the swelling time.

\end{abstract}

\maketitle

\section*{Introduction}

Swelling of polymer gels is one of the classical problems in polymer science and technology. Many important ramifications for technical applications, ranging from simulate biological tissues to controlled drug release to gel actuators, are directly related to the gel swelling~\cite{Okuzaki19}. 

A dried gel placed in a solvent swells absorbing solvent from the surrounding. This process is caused by the thermodynamic mixing energy between polymer and solvent. One may therefore understand the swelling process as a diffusion process of solvent in polymer network. Though such picture captures an essential aspect of the phenomena, the diffusion in gels is different from the diffusion in solutions. The difference arises from the fact that gel is an elastic material, and therefore the solvent diffusion is coupled with the deformation of the polymer network, and is coupled with the stress field created in the gel.  Such coupling is called diffusio-mechanical coupling~\cite{Doi09, Doi21}. 

The effect of diffusio-mechanical coupling was well aware of at an early study of gel physics.  It was realized that the collective diffusion constant of a gel depends on the shear modulus $G$~\cite{TanakaBen}, and this diffusion constant has to be used in the swelling time of a gel~\cite{TanakaFillmore}. On the other hand, the effect was discussed in an approximate manner, and it took some time for the full set of equations to be completed. 

Indeed the theory of swelling dynamics of gels has some history.  The first theory was given by Tanaka and Filmore in 1979~\cite{TanakaFillmore}. They proposed an equation (TF equation) which accounts for the effect of elasticity by two material constants, the osmotic bulk modulus $K$ and the shear modulus $G$. They solved the TF equation for spherical gel assuming that $G/K$ is negligibly small.   The TF equation was solved by Peters and Candau~\cite{Peters86, Peters88} for cylinder and disk-shaped gels, and for non-negligible values of $G/K$. However, in 1989, Li and Tanaka~\cite{LiTanaka} pointed out a fundamental deficit of  the TF equation: the swelling time of a cylindrical gel in radial direction and in axial direction are different by a factor of the square of the aspect ratio (the ratio of length and radius). To remove the deficit, they realized that the relaxation of polymer network needs to be accounted , and they proposed an approximate way of dealing with it. Wang et al ~\cite{Hu97} gave an improved theory, but it still involves an approximation which is valid for dilute polymer concentration. A complete set of equations describing the diffusio-mechanical coupling was given in 1992~\cite{DoiOnuki92}, and the solution of the equation for the cylindrical gel was obtained in 2005~\cite{Doi05}. To date, a rigorous solution for the swelling of a disk-shaped gel is still missing.

In this paper, we shall discuss the swelling dynamics of a disk-shaped, or more generally a sheet-like, gel. This work was motivated by two reasons. (a) The gel dynamics is now being quantitatively studied for precisely controlled gel samples~\cite{Sakai19,Sakai20a,Sakai20b}, and the effect of shear modulus $G$ has been studied.  Results of rigorous calculation is needed in such studies.  We will show that the characteristic features of the swelling of a disk gel are not revealed by previous theories. (b) Though the diffusio-mechanical equations can be solved analytically here, they are usually difficult to solve. We shall study how an approximate calculation based on the variational principle works well in present problem.

In the first part of this paper, we shall solve the diffusio-mechanical equation analytically for disk-shaped gel, and show some unusual consequences derived by the equation.  We shall show that although the macroscopic aspect ratio of the gel remains unchanged during the swelling, the aspect ratio of the internal element changes non-uniformly; the element near the mid plane is compressed in axial direction, and stretched in radial direction. We shall also show that it is wrong to say that the swelling time is dominated by the collective diffusion constant since the swelling time diverges in the limit that the osmotic bulk modulus $K$ goes to zero even though the collective diffusion constant remains finite. 

In the second part of this paper, we shall solve the problem approximately using the variational principle which underlies the diffusio-mechanical equation.  We shall show that all the above unusual features are captured by the approximate solution obtained by this method. An analytical expression is given for the swelling time by the variational method.

\begin{figure*}[h!t]
{\includegraphics[width=0.75\textwidth,draft=false]{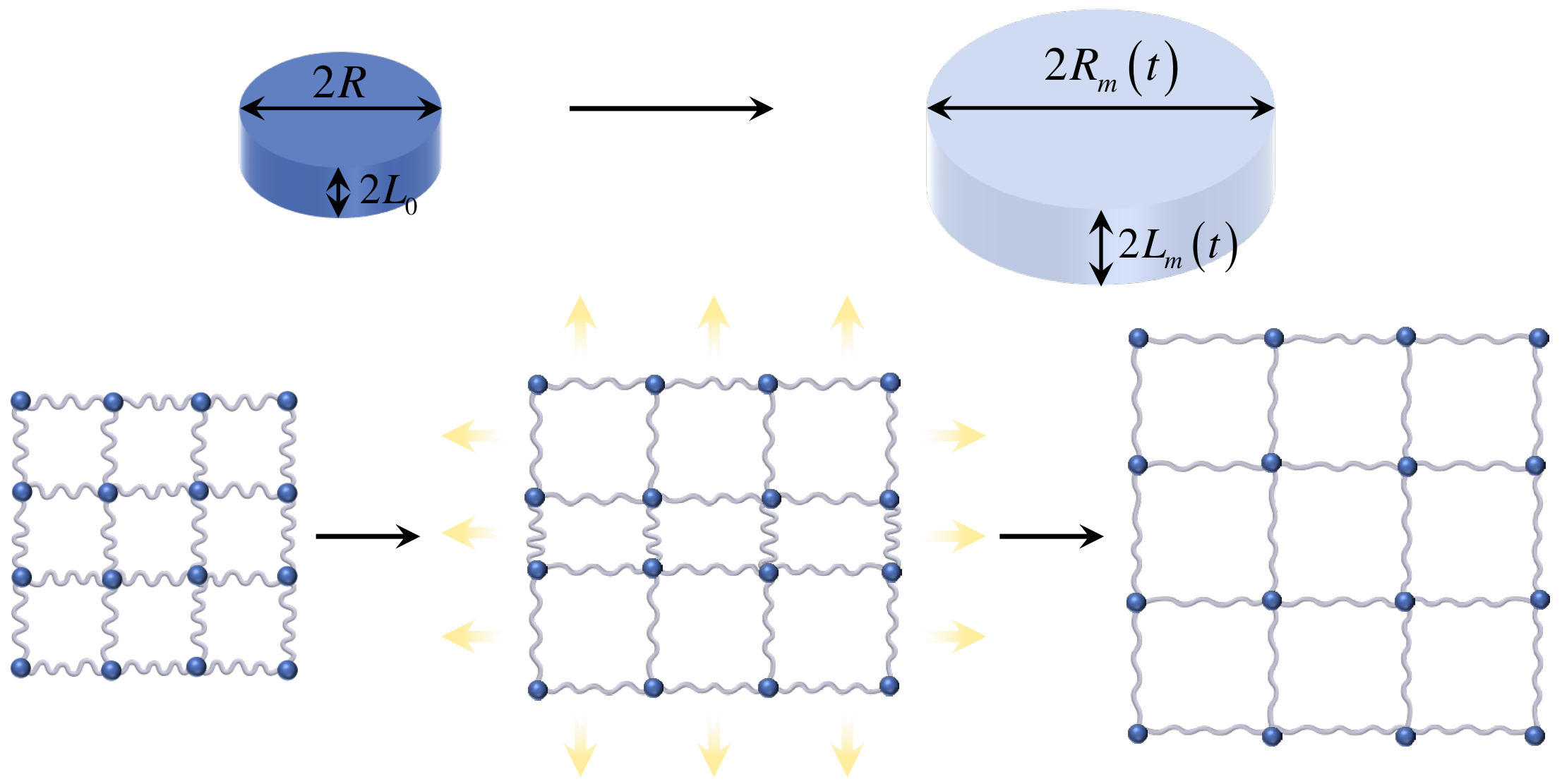}}
\caption{
\textsf{Top panel: a gel initially having radius $R$ and thickness $2L_0$, swells to have radius $R_{m}(t)$ and thickness $2L_m(t)$ at time $t$. Bottom panel:a  phenomenon seen at the early swelling stage. The gel is compressed near the mid-plane while its radius and thickness increases monotonically. The gel eventually swells to an equilibrium state.}}
\label{fig1}
\end{figure*}

\section*{Theoretical Framework}
We consider a disk-shaped gel which was initially at an equilibrium reference state with radius $R$ and half thickness $L_0$, as shown in Fig.~1. We choose the cylindrical coordinate system such that the top and bottom surfaces of the disk are at $\pm L_0$, respectively. The origin $z=0$ corresponds to the plane of symmetry of the disk. 

When such a gel is placed in a solution, the gel absorbs solvent and starts to swell toward new equilibrium state. Since the disk is thin, the deformation is uniform in the radial direction, but is non-uniform in the axial direction. Noticing that the deformation has a reflection symmetry with respect to the $z=0$ plane. The point on the gel network which was located at $(r,z)$ in the reference state is displaced to $[(1+\epsilon(t))r,z+u(z,t)]$ at time $t$, where $u(z,t)$ is the displacement in $z$ direction. We here only consider the cases that the deformation from the reference state is small. We assume that $\epsilon(t) \ll 1$ and $u(z,t)\ll L_0$. The time evolution of $\epsilon(t)$ and $u(z,t)$ are determined by the Onsager variational principle. We show here two methods to calculate the swelling dynamics of a disk-shaped gel, including the diffusio-mechanical equations and an approximate variational method.

\subsection*{The diffusio-mechanical equations}
The Onsager variational principle states that the evolution of the system is determined by the minimum of the Rayleighian function defined by
\begin{equation}\label{Ray}
\Re=\Phi+\dot{F}
\end{equation}
where $\dot F$ is the time derivative of the free energy, and $\Phi$ is the energy dissipation function of the system.

The deformation free energy of the gel has been discussed in the textbook of polymer physics~\cite{Doi13}.  For uniaxial deformation, it is written as
\begin{equation}\label{fre}
\frac{F}{\pi R^2}=\int^{L_0}_{-L_0}dz\left[\frac{K}{2}\left(2\epsilon+\frac{\partial u}{\partial z}-3\epsilon_{\rm{eq}}\right)^2+
\frac{2G}{3}\left(\epsilon-\frac{\partial u}{\partial z}\right)^2\right]
\end{equation}
where $\epsilon_{\rm{eq}}$ represents the equilibium strain toward which $\epsilon(t)$ and 
$\partial u/\partial z$ change in time, and $K$ and $G$ are material constants, called osmotic bulk modulus and shear modulus respectively.

The time derivative of such a free energy is easily obtained as
\begin{equation}\label{dotF}
\begin{aligned}
\frac{\dot{F}}{\pi R^2}=&\left[K\left(2\epsilon+\frac{\partial u}{\partial z}-3\epsilon_{\rm{eq}}\right)-
\frac{4G}{3}\left(\epsilon-\frac{\partial u}{\partial z}\right)\right]\dot{u}\bigg|^{L_0}_{-L_0}+\\
&\int^{L_0}_{-L_0}\left\{ -\dot{u}\left(K+\frac{4G}{3}\right)\frac{\partial^2 u}{\partial z^2}+
 \dot{\epsilon}\left[\left(4K+\frac{4G}{3}\right)\epsilon+\left(2K-\frac{4G}{3}\right)\frac{\partial u}{\partial z}-6K\epsilon_{\rm{eq}}\right] \right \}
\end{aligned}
\end{equation}

The energy dissipation of the gel is caused by the motion of polymer relative to that of solvent.  Let $\bm{V}_p$ and $\bm{V}_s$ be the velocity vector of polymer and that of solvent at point $(r,z)$, then  the energy dissipation is given by
\begin{equation}\label{phi0}
   \Phi= \int_0^{R}  2\pi r dr \int_{-L_0}^{L_0} dz  \frac{\xi}{2} \left(1+\frac{\partial u}{\partial z}\right)\left(1+\epsilon\right)^2
                            (\bm{V}_p - \bm{V}_s)^2
\end{equation}
where $\xi$ is the friction constant per unit volume for the polymer motion relative to solvent.

The radial and axial component of $\bm{V}_p$ is given by 
$V_{\rm{pr}}=\dot \epsilon r$, and $V_{\rm{pz}}= \dot u$.  Notice that $V_{\rm{pr}}$ is much 
larger that $V_{\rm{pz}}$ for thin disk shaped gel of $R \gg L_0$. This means that the 
relative motion between polymer and solvent in radial direction will not take place since
such motion causes a very large energy dissipation.  Therefore, we may assume 
$V_{\rm{pr}}=V_{\rm{sr}}$, and the relative motion takes place only in $z$-direction.  Hence
eq.(\ref{phi0}) can be written as
\begin{equation}
   \frac{\Phi}{\pi R^2}= \int_{-L_0}^{L_0} dz \frac{\xi}{2}
                            (V_{\rm pz} - V_{\rm sz})^2
\end{equation}
where we have ignored the small terms 
$\partial u/ \partial z$ and $\epsilon^2$ which appear in eq.(\ref{phi0}).

The velocity $\bm{V}_p$ and  $\bm{V}_s$ have to satisfy the incompressible condition $\nabla\cdot[\phi \bm{V}_{p}+(1-\phi) \bm{V}_{s}]=0$. Due to the convention that $V_{\rm pr}=V_{\rm sr}=\dot{\epsilon} r$, the incompressible condition becomes
\begin{equation}
\frac{\partial [\phi V_{\rm pz}+(1-\phi) V_{\rm sz}]}{\partial z}+2\dot{\epsilon}=0
\end{equation}
Then, the axial component of $\bm{V}_s$ is written as
\begin{equation}
V_{\rm sz}=\frac{-2z\dot{\epsilon}-\phi\dot{u}}{1-\phi}
\end{equation}
Inserting the expression of $V_{\rm{pz}}$ and $V_{\rm{sz}}$ into $\Phi$, the dissipation function becomes
\begin{equation}\label{phi}
\frac{\Phi}{\pi R^2}=\int^{L_0}_{-L_0}dz \frac{ \tilde{\xi} }{2}\left(\dot{u}+2z\dot{\epsilon}\right)^2
\end{equation}
where $\tilde{\xi}=\xi/(1-\phi)^2$ is defined as the effective friction constant. 

The Onsager principle states that the time evolution of $u(z,t)$ and $\epsilon(t)$, are determined by the condition $\partial \Re/\partial \dot{u}=0$ and $\partial \Re/\partial \dot{\epsilon}=0$, respectively. Since the present system has a 
reflection symmetry with respect to the mid-plane at $z=0$, we consider the 
 the half part of $z\in[0,L_0]$. By 
eqs. (\ref{Ray}), (\ref{dotF}) and (\ref{phi}), this gives the following evolution equations
\begin{equation}\label{dotu}
\tilde{\xi}(\dot{u}+2z\dot{\epsilon})-\left(K+\frac{4G}{3}\right)\frac{\partial^2 u}{\partial z^2}=0
\end{equation}
\begin{equation}\label{dotep}
\int^{L_0}_{0}dz\tilde{\xi}(\dot{u}+2z\dot{\epsilon})2z+
\int^{L_0}_{0}dz\left[\left(4K+\frac{4G}{3}\right)\epsilon+\left(2K-\frac{4G}{3}\right)\frac{\partial u}{\partial z}-6K\epsilon_{\rm eq}\right]=0
\end{equation}
and the boundary conditions
\begin{eqnarray}
u(0,t)&=&0,       \label{bdary1} \\
\left(K+\frac{4G}{3}\right)\frac{\partial u}{\partial z}\bigg|_{z=L_0}&=&\left(\frac{4G}{3}-2K\right)\epsilon+3K\epsilon_{\rm eq}      \label{bdary2}
\end{eqnarray}
Notice that the boundary condition at $z=0$, eq. (\ref{bdary1}), is due to the  
symmetry with respect to the mid-plane: the position of the mid-plane is fixed for all time.

Inserting eq. (\ref{dotu}) and the boundary conditions (\ref{bdary1}), (\ref{bdary2})
 into eq. (\ref{dotep}), we have
\begin{equation}\label{epsi}
\epsilon(t)=\frac{u(L_0,t)}{L_0}
\end{equation}
Using the radius $R_m(t)$, and the thickness $2L_m(t)$ of the gel (see Fig.1), eq. (13) is written as $[R_m(t)-R_m(0)]/R_m(0)=[L_m(t)-L_m(0)]/L_m(0)]$, or $R_m(t)/R_m(0)=L_m(t)/L_m(0)$. This indicates that the aspect ratio of the gel $R_m(t)/L_m(t)$ remains unchanged during swelling: the gel looks to swell isotropically.  Such isotropic swelling is apparent since individual element of the gel do not swell isotropically: they are deformed anisotropically during the swelling process as we shall show later. 

Defining the collective diffusion constant $D_{\rm{eff}}=\left(K+\frac{4G}{3}\right)/\tilde{\xi}$ as in the theory of Tanaka and Fillmore~\cite{TanakaFillmore}, eq. (\ref{dotu}) becomes
\begin{equation}\label{dfu}
\dot{u}=D_{\rm{eff}}\frac{\partial ^2 u}{\partial z^2}-2z\dot{\epsilon}
\end{equation}
where the derived $D_{\rm{eff}}$ as a function of $K$, $G$ and $\tilde{\xi}$ is consistent with the classical theory~\cite{TanakaBen}, and has been quantitatively validated by recent experiments~\cite{Sakai19, Sakai20a} that Sakai's group measured the diffusion coefficient and the shear modulus of Tetra-poly(ethylene glycol) gels by dynamic light scattering and dynamic viscoelastic measurement.

Finally, eqs. (\ref{epsi}) and (\ref{dfu}) together with the boundary condition eqs. (\ref{bdary1}) and (\ref{bdary2}), and the initial conditions $\epsilon(0)=0$ and $u(z,0)=0$ are the set of the diffusio-mechanical equations of the swelling of disk-shaped gel. Solving such set of equations is not trivial.  If $\epsilon(t)$ is known, eq. (\ref{dfu}) can be solved easily for $u(z,t)$.  However, in the present problem, $\epsilon(t)$ is an unknown which has to be solved simultaneously together with $u(z,t)$ using the boundary condition eq.~(\ref{bdary2}).  This is a characteristic of the diffusio-mechanical coupling problem. Equation (\ref{dfu}) indicates that the displacement in $z$ direction is coupled with the displacement in radial direction.  Such coupling arises from the mechanical coupling of polymer network in the gel.  Though the structure of the equations are slightly involved, the set of equations are linear, and can be solved exactly.  This is shown in the next section.

\subsection*{Exact solution}

To solve the set of evolution equations, we define $g(z,t)=u(z,t)+2z\epsilon-3z\epsilon_{\rm{eq}}$. Then, eq. (\ref{dfu}) and the boundary conditions become
\begin{eqnarray}\label{dfug}
\frac{\partial g}{\partial t}&=&D_{\rm{eff}}\frac{\partial ^2 g}{\partial z^2},\\
g(0,t)&=&0,     \\
\frac{\partial g}{\partial z}\bigg|_{z=L_0}&=&\beta \frac{g(L_0,t)}{L_0}
\end{eqnarray}
where $\beta=4G/(4G+3K)$.

The above set of equations can be solved by the standard method of eigen function expansion. Taking $L_0$, as the unit of length, the solution can be written as
\begin{equation}\label{solg}
g(z,t)=\sum^{+\infty}_{n=1}C_{n}e^{-D_{\rm{eff}} \lambda^2_n t} 
                             \sin\left(\lambda_n z\right),
\end{equation}
and the expansion coefficients are
\begin{equation}
C_n=\frac{\int^1_0 g_0(z)\sin(\lambda_n z) dz}{\int^1_0\sin^2(\lambda_n z) dz}
\end{equation}
where $g_0(z)=-3z\epsilon_{\rm{eq}}$, and $\lambda_n$ is the $n$-th solution of the equation
\begin{equation}\label{lam1}
\tan \lambda =\frac{\lambda}{\beta}
\end{equation}
Here, we only consider positive solutions, i.e. $\lambda_n>0$.

The volume strain (the volume change per volume) is
defined by $w(z,t)=2\epsilon(t)+\frac{\partial u}{\partial z}$, and  is given by
\begin{equation}\label{vols}
w(z,t)=\sum^{+\infty}_{n=1}C_{n} \lambda_n e^{-D_{\rm{eff}} \lambda^2_n t} \cos(\lambda_n z)+3\epsilon_{\rm{eq}},
\end{equation}
%

\subsection*{Approximate solution by variational method}

Though the diffusio-mechanical equations can be solved analytically, the solution is not simple and requires some numerical calculation in getting the final result. In this subsection, we show a different strategy to solve the problem: the approximate method based on the same Onsager variational principle.  In this method, we reduce the partial differential equations to ordinary differential equations.  This method has been successfully used in the evaporation of liquid droplets~\cite{Man16,Man17, Man20, Man21}. We shall show that in the present problem, the main feature of the exact solution is captured by the variational method.

In the variational method, we follow the same derivation process as the exact model, but assume an approximate form of the displacement vector in the $z$-direction by $u(z,t)=\beta_1(t)z+\beta_2(t)z^2$. Then, the deformation of the gel is now described by $\epsilon$, $\beta_1$, and $\beta_2$ in the variational method instead of $\epsilon$ and $u(z)$ in the exact solution.

Under such convention, the free energy eq. (\ref{fre}) now becomes
\begin{equation}\label{vfre}
\frac{F}{\pi R^2}=\int^1_0 dz \left[\frac{K}{2}\left(2\epsilon+\beta_1+2z\beta_2-3\epsilon_{eq}\right)^2+
\frac{2G}{3} \left(\epsilon-\beta_1-2z\beta_2 \right)^2 \right]
\end{equation}
where we also set $L_0$ as the unit of length. The integral in eq.~(\ref{vfre}) can be
calculated easily, and its time derivative is calculated as
\begin{equation}\label{dvfre}
\begin{aligned}
\frac{\dot{F}}{\pi R^2}=&\left[4K\epsilon+2K\beta_1-6K\epsilon_{eq}+2K\beta_2+\frac{4G}{3}(\epsilon-\beta_1-\beta_2)\right]\dot{\epsilon}+\\
&\left[2K\epsilon+K\beta_1-3K\epsilon_{eq}+K\beta_2-\frac{4G}{3}(\epsilon-\beta_1-\beta_2)\right]\dot{\beta}_1+\\
&\left[2K\epsilon+K\beta_1-3K\epsilon_{eq}+\frac{4}{3}K\beta_2-\frac{4G}{3}(\epsilon-\beta_1-\beta_2)+\frac{4}{9}G\beta_2\right]\dot{\beta}_2
\end{aligned}
\end{equation}

The velocity of polymer in $z$-direction becomes $V_{\rm{pz}}=\dot{\beta}_1 z+\dot{\beta}_2 z^2$. Therefore, the dissipation function eq. (\ref{phi}) is now written as
\begin{equation}\label{vphi}
\frac{\Phi}{\pi R^2}=\int^1_0 dz \frac{ \tilde{\xi} }{2}\left(\dot{\beta}_1 z+\dot{\beta}_2 z^2+2z\dot{\epsilon}\right)^2
\end{equation}

Equations (\ref{dvfre}) and (\ref{vphi}) construct a new Rayleighian. Then, following the same minimization process as in the previous subsection with respect to $\dot{\epsilon}$, $\dot{\beta}_1$ and $\dot{\beta}_2$, we have the following set of evolution equations
\begin{equation}\label{vset}
\begin{aligned}
\title{\xi}\dot{\beta}_1&=52K(\beta_1-\epsilon_{eq})+\left(\frac{688}{9}K+\frac{880}{27}G\right)\beta_2,\\
\title{\xi}\dot{\beta}_2&=-60K(\beta_1-\epsilon_{eq})-\left(\frac{260}{3}K+\frac{320}{9}G\right)\beta_2,\\
\dot{\epsilon}&=\dot{\beta}_1+\dot{\beta}_2
\end{aligned}
\end{equation}
The above set of equations can be analytically solved. Detailed derivation of the solution can be found in the supplementary material. The solutions are written as
\begin{equation}\label{vsol}
\begin{aligned}
\beta_1&=c_1e^{-\lambda_{-}t}+c_2e^{-\lambda_{+}t}+\epsilon_{eq},\\
\beta_2&=c_1a_{-}e^{-\lambda_{-}t}+c_2a_{+}e^{-\lambda_{+}t}
\end{aligned}
\end{equation}
where  $\lambda_{\pm}=\frac{156K+160G \pm 8\sqrt{279K^2+645GK+400G^2}}{9\tilde{\xi}}$, $a_{\pm}=-\frac{78K+20G\pm\sqrt{279K^2+645GK+400G^2}}{86K+\frac{110}{3}G}$, $c_1=-\frac{\epsilon_{eq}a_{+}}{a_{+}-a_{-}}$ and $c_2=\frac{\epsilon_{eq}a_{-}}{a_{+}-a_{-}}$, which are all constants related to the material parameters $K$ and $G$.

\section*{Results and Discussions}

We now show the above results in graphical representation. Unless otherwise specified, we take $L_0$ as the unit of length and the unit of time scale by the conventional  characteristic time $\tau_{\rm TF}$ defined by
\begin{equation}\label{stem0}
    \tau_{\rm TF}=\frac{L^2_0}{D_{\rm{eff}}}= \frac{L^2_0 \tilde{\xi}}{ K+\frac{4G}{3} }
\end{equation}
In this convention, we have equalities $D_{\rm{eff}}=1$, and $\tilde{\xi}=K+\frac{4G}{3}$.  Therefore there is only one dimensionless parameter, $G/K$,to be studied.  Notice that  the parameter $\epsilon_{\rm{eq}}$ is irrelevant in our analysis of linear set of equations.

\begin{figure*}[h!t]
{\includegraphics[width=0.95\textwidth,draft=false]{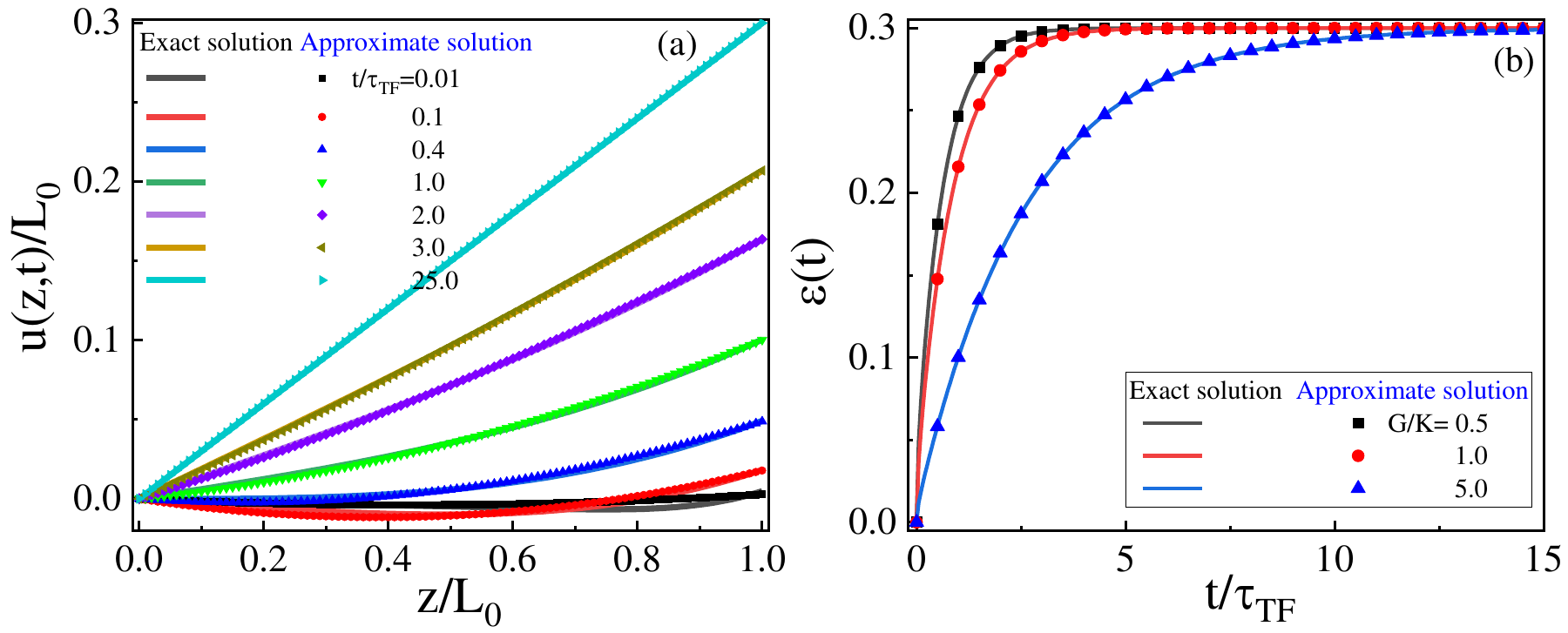}}
\caption{
\textsf{The comparison of the relative displacements obtained from the exact solution (solid lines) and from the approximate solution (data dots). (a) the time evolution of axial direction displacement vector $u(z,t)/L_0$ for $G/K=5.0$; (b) the time evolution of the radial direction displacement vector $\epsilon(t)$ for three values of $G/K=0.5$, $1.0$, and $5.0$. For all calculations $\epsilon_{\rm{eq}}=0.3$, and the time scale is $\tau_{\rm TF}=L^2_0/D_{\rm{eff}}$. 
}}
\label{fig2}
\end{figure*}

\subsection*{ Time evolution }

A comparison of the displacements obtained from the exact solution and the variational method are shown in Fig.~2. Figure~2a shows $u(z,t)$, where all solid lines are calculated by eq.~(\ref{solg}) of the exact solution and all data points are obtained from $u(z,t)=\beta_1 z+\beta_2 z^2$ with $\beta_1$ and $\beta_2$ calculated by eq.~(\ref{vsol}). Results shows good agreement between the two models. Interestingly, figure 2a indicates that in the very beginning stage, the element near the surface of the gel is stretched in $z$-direction ($\partial u/\partial z > 0$) while the elements near the mid-plane is compressed ($\partial u/\partial z < 0$). Eventually, all gel elements become stretched in $z$ direction ($\partial u/\partial z = \epsilon_{eq}$ for $t \to \infty$). Figure 2b is the relative displacement $\epsilon(t)$ in the radial direction for three values of $G/K=0.5$, $1.0$, and $5.0$. Good agreement between the exact solution and the variational solution is also seen. Results show that the radius of the gel increases monotonically in time, and finally stops increasing at the equilibrium value $\epsilon_{\rm{eq}}$. For a fixed $\epsilon_{\rm{eq}}$, the swelling time is longer for larger value of $G/K$.

\begin{figure}[h!t]
{\includegraphics[width=0.9\textwidth,draft=false]{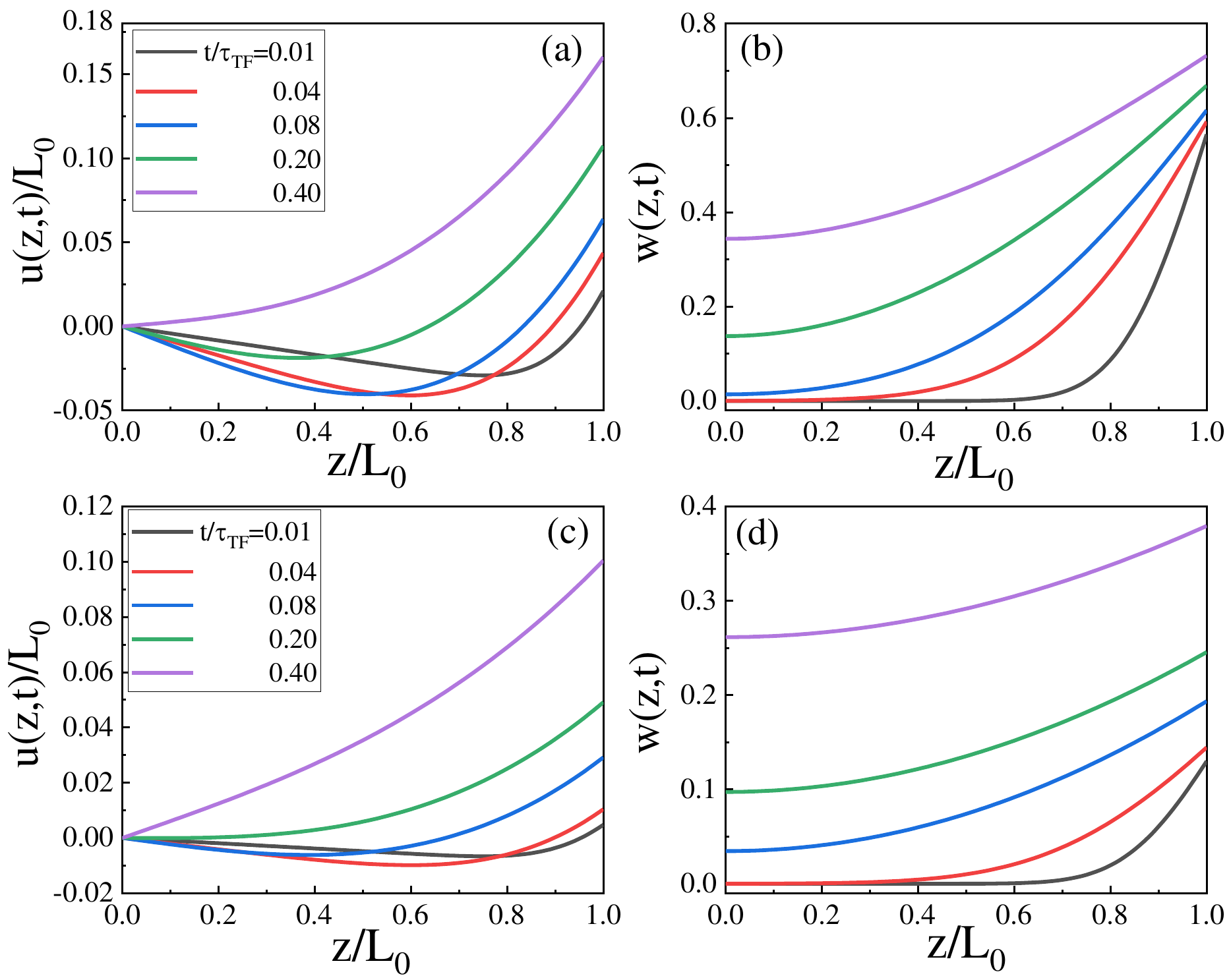}}
\caption{
\textsf{The time evolution of $u(z,t)/L_0$ and the volume strain in the early swelling stage calculated from the exact solution for $G/K=0.5$ in (a) and (b), and $G/K=5.0$ in (c) and (d). While the gel thickness increases monotonically, the gel is compressed in the bulk part near the mid-plane. The volume strains $w(z,t)$ are always positive, i.e., $w(z,t)\ge 0$, for all time. For all calculations, $\epsilon_{\rm{eq}}=0.3$ and the time scale is $\tau_{\rm TF}=L^2_0/D_{\rm{eff}}$. 
}}
\label{fig3}
\end{figure}

The compression during the swelling process has not been reported previously. We therefore zoomed into the early stage, and studied this abnormal phenomenon. Figure 3a and c show the time evolution of $u(z,t)$. It is seen that at an early stage, the displacement $u(z,t)$  becomes negative for small $z$.  This means that while the gel as a whole swells, the bulk elements are compressed.  The reason for the compression is seen by looking at the time evolution of the volume strtain $w(z,t)$ (see Figure 3b and d).  When the swelling starts, $w(z,t)$ increases quickly near the surface, while it remains to be zero near the midplane.   This is quite natural since  solvent first swells the gel surface and penetrates into the bulk gel.  Therefore at the early stage of swelling, the surface elemenets are swollen, while the inner elemements are unswollen.  The swollen surface elements stretches the disk in radial direction and gives the monotonically increasing radial strain $\epsilon(t)$ shown in Fig.~2. The positive radial strain causes the compression in axial direction for the unswollen element near the mid-plane.  

The degree of compression depends on the value of the shear modulus $G$. If $G$ is very large, the axial strain $\partial u/\partial z$ has to be equal to $\epsilon$ (see eq.(\ref{fre})), and there will no compression.  We therefore expect that the degree of the compression decreases with the increase of $G/K$.  Such tendency is indeed seen in Figure 3. It is seen that the compression is more prominent for $G/K=0.5$ (Fig.~3a) than for $G/K=5.0$ (Fig.~3c). This phenomenon can be checked by future experiments.

\subsection*{ Swelling time }  

The swelling time can be defined as the longest relaxation time in the time evolution of 
$u(z,t)$ or $\epsilon(t)$.  In the exact solution, this is given by
\begin{equation}\label{stem}
    \tau=\frac{L^2_0}{\lambda^2_1 D_{\rm{eff}}} =\frac{L^2_0 \tilde{\xi} }{ K+\frac{4G}{3} }
\end{equation}
where $\lambda_1$ is the smallest positive solution of eq.~(\ref{lam1}).
In the variational calculation, $\tau$ is given by $L^2_0/\lambda_{-}$.
The two swelling times obtained from exact solution and variational calculation are plotted in Fig. 4a as a function of $G/K$.
Results show good agreement between the two models, and $\tau/\tau_{\rm TF}$ is an increasing function of $G/K$. Such result has
been reported in the previous studies\cite{LiTanaka,Hu97,Doi05}.

\begin{figure*}[h!t]
{\includegraphics[width=0.95\textwidth,draft=false]{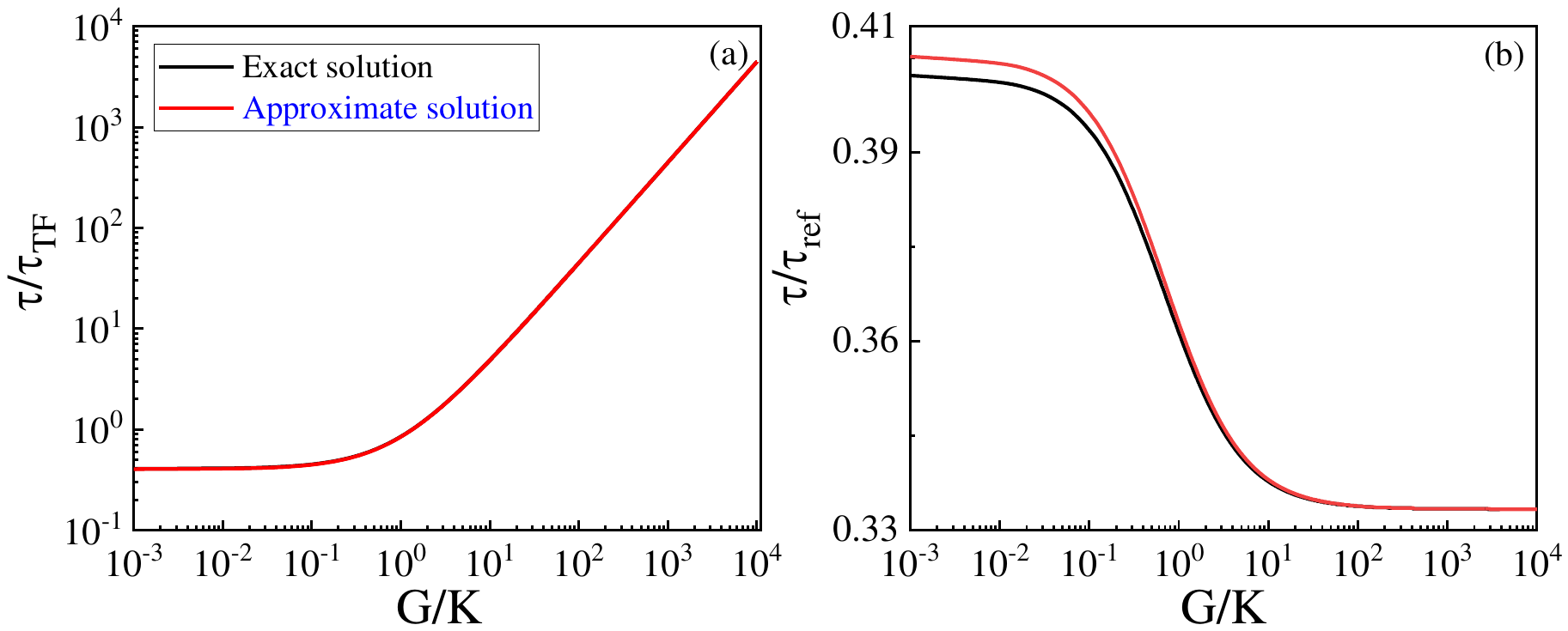}}
\caption{
\textsf{The dependence of the longest swelling time on $G/K$: (a) rescaled by $\tau_{\rm TF}$, (b) rescaled by $\tau_{\rm ref}$. The black lines are obtained by the exact solution, while the red lines are the approximate solution. In the limit of $G/K\to\infty$,  $\tau/\tau_{\rm TF} \to \infty$ and $\tau/\tau_{\rm{ref}} \to 1/3$ for both solutions. On the other hand, in the limit of $G/K\to0$, both $\tau/\tau_{\rm TF}$ and $\tau/\tau_{\rm{ref}}$ approach to $0.402$ for exact solution, while they approach to $0.405$ for the approximate solution.  
 }}
\label{fig4}
\end{figure*}

In our theory, 
analytical expression for the swelling time  can be obtained for the two limits,
$G/K \ll 1$ and  $G/K \gg 1$. For the exact solution, they are given by
\begin{eqnarray} 
   &\mbox{for}& \quad G/K \ll 1  \qquad   \tau/\tau_{\rm TF} = 4/\pi^2\approx 0.405    \\
   &\mbox{for}& \quad  G/K \gg 1 \qquad   \tau/\tau_{\rm TF} = 4G/9K  
\end{eqnarray}
For the variational solution
\begin{eqnarray} 
 &\mbox{for}& \quad G/K  \ll 1 \qquad   \tau/\tau_{\rm TF}  \approx  0.402  \\
 &\mbox{for}& \quad  G/K \gg 1 \qquad   \tau/\tau_{\rm TF}  =  4G/9K  
\end{eqnarray}
These results indicate clearly that $\tau/\tau_{\rm TF}$ increases indefinitely
as the shear modulud $G$ increases. However, it must be remembered
that $\tau_{\rm TF}$ is not constant when $G$ is changed.  

In fact, if we choose 
a reference time which is independent of $G$, the swelling time does not vary so much 
with $G$. Figure 4b shows such a plot, where the reference time $\tau_{\rm ref}$ is defined by
\begin{equation}  \label{tau_F}
  \tau_{\rm{ref}}=\frac{L^2_0\tilde{\xi}}{K}
\end{equation}
which is independent of $G$. It is seen $\tau/ \tau_{\rm{ref}}$ remains in the range between 0.33 and 0.41.  
It is therefore misleading to say that the swelling time is determined by
the collective diffusion constant $D_{\rm eff}$. Swelling time is essentially estimated by
the osmotic bulk modulus (i.e., $\tau \simeq L_0^2 \tilde{\xi}/K$), and the 
effect of shear moduls is small on the swelling time. In fact, it has been proven 
that for any shape of the gel, the swelling time (i.e., the longest relaxation
time)  is proportional to $1/K$ in the two limits of $G/K \to 0$ and $G/K \to \infty$. 
At this moment, it is a theoretical prediction. It will be interesting to  check the prediction by experiments since this feature of the swelling time does not seem to be widely known in gel community. 

\section*{Conclusions}
We have presented a theory for the swelling of a disk-shaped gel in solution. The equation accounts for the elastic deformation of polymer network, and the diffusive motion of solvent in polymer network. The theory provides explicit expressions of the time evolution of the displacements in both radial and axial direction. The theory shows a novel phenomenon that while the thickness and the radius of the gel increases monotonically in time with keeping its original shape during the whole swelling process, the gel is 
compressed in axial direction near the mid-plane at the early swelling stage. An analytical expression is given for the swelling time $\tau$, indicating that $\tau$ is essentially determined by osmotic bulk modulus $K$ and weakly dependent on shear modulus $G$, and  
decreases when $G$ is increased. These results can be checked by future experiments. 

We also proposed a new strategy to solve the diffusio-mechanical problem.  It is based on the variational principle which underlies the diffusio-mechanical equation. While this method is not exact, it gives an explicit analytical solution that is in good agreement with the exact solution.

Our theory predicts that for small deformation, disk gel shows isotropic swelling. This feature has been reported long time ago by Li and Tanaka, where they experimentally measured the swelling of a long cylinder and disk acrylamide gel~\cite{LiTanaka}. Isotropic swelling of cylinder and disk gel was also indicated by previous theory~\cite{LiTanaka, Hu97, Doi05}. About the phenomenon of compression during a swelling process, however, there is no report as far as we know, and it would be interesting to observe it experimentally. Our analysis of the swelling time is qualitatively consistent with previous theories as long as the conventional characteristic time $\tau_{\rm TF}$ is used as the reference time, (see  Fig. 4a), but we would like to emphasize that if a different reference time is used, the swelling time becomes almost independent of G, leading a new dependence behavior of the swelling time on $G/K$.
 
We hope that this kind of analytical work provides some insight into the design and applications of polymer gel materials.

\bigskip
{\bf Acknowledgement.}~~
X. M. thanks Y. Wang and C. H. Huang for useful discussions. This work was supported in part by grants No.~21822302 of the National Natural Science Foundation of China (NSFC), and the NSFC-ISF Research Program, jointly funded by the NSFC under grant No.~21961142020, and the Israel Science Foundation (ISF) under grant No.~3396/19.

{\bf Supporting Information}~~
Detailed derivations of the model by a variational method

\newpage

\clearpage
\vskip 0.5truecm
\centerline{for Table of Contents use only}
\centerline{\bf Swelling Dynamics of a Disk-Shaped Gel}
\centerline{\it Xingkun Man*, and Masao Doi*}

\begin{figure}[h]
{\includegraphics[width=0.7\textwidth,draft=false]{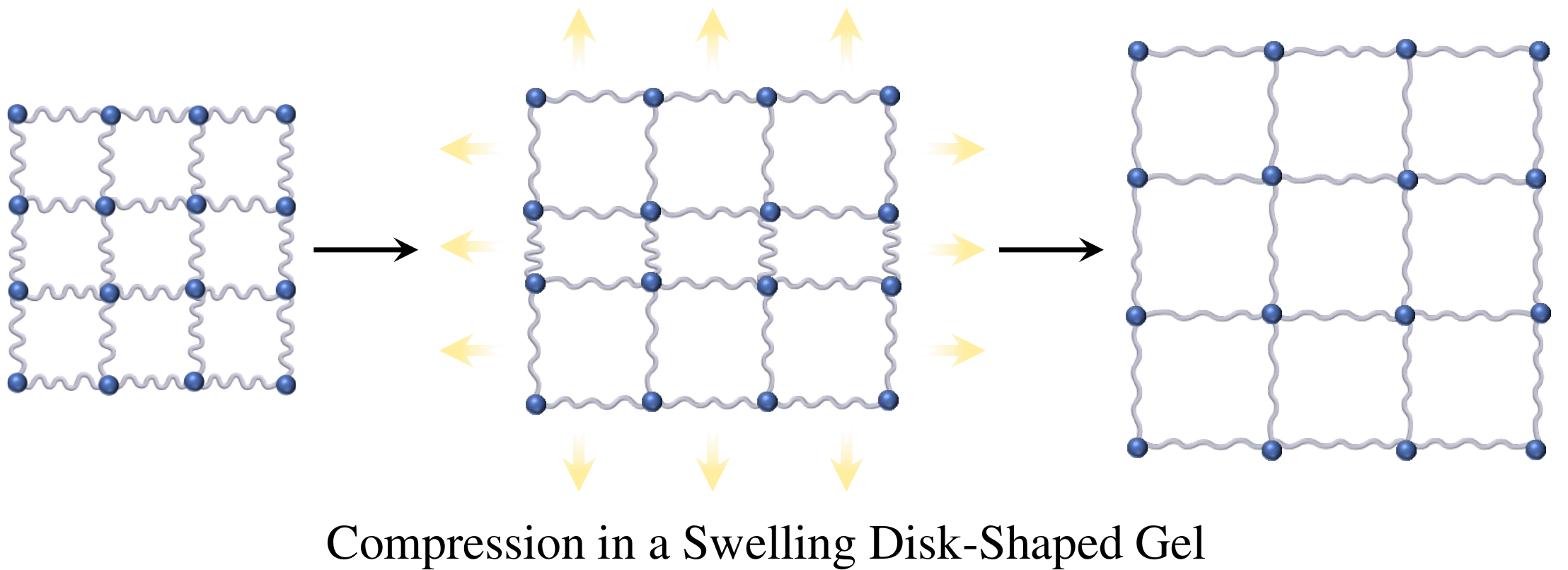}}
\end{figure}

\end{document}